%% file: paper.tex
\documentclass[sigconf,natbib=false]{acmart}

\usepackage{orcidlink}
\usepackage{booktabs}
\usepackage{tabularx}
\usepackage{float}
\usepackage{caption}
\usepackage{subcaption}
\usepackage{listings}
\usepackage[ruled]{algorithm2e}
\usepackage{float}
\usepackage{varioref}
\usepackage[capitalise]{cleveref}
\usepackage{enumitem}
\usepackage[dvipsnames]{xcolor}

\usepackage{amsthm}
\theoremstyle{definition}
\newtheorem{definition}{Definition}[section]

\usepackage{epsf,picinpar}
\usepackage[color=red!50!white,textsize=scriptsize,textwidth=3.5cm]{todonotes}
\usepackage{soul}
\newcommand{\secref}[1]{Sec.~\ref{#1}}
\newcommand{\tabref}[1]{Tab.~\ref{#1}}
\newcommand{\figref}[1]{Fig.~\ref{#1}}

\newcommand{\listref}[1]{Lst.~\ref{#1}}

\definecolor{dslblue}{RGB}{0,70,150}
\definecolor{dslpurple}{RGB}{130,0,150}
\definecolor{dslgray}{RGB}{100,100,100}
\definecolor{dslgreen}{RGB}{0,120,50}
\definecolor{dslred}{RGB}{180,0,0}
\lstdefinelanguage{ContractDSL}{
  morekeywords=[1]{service, inputs, outputs, assumes, guarantees, over},
  morekeywords=[2]{FRESHNESS, COMPLETENESS, ACCURACY, AVAILABILITY,
                   SAMPLING_RATE, MAX_SKEW, CONSISTENCY, GRANULARITY},
  morekeywords=[3]{<=, >=, ==},
  sensitive=true,
  morecomment=[l]{//},
  morestring=[b]"
}
\input{preprint-commands}

\newcommand{\TODO}[1]{
    \ifthenelse{\equal{#1}{}}{\textcolor{VioletRed}{TODO}}{\textcolor{VioletRed}{TODO:~{#1}}}%
}

\makeatletter
 \if@todonotes@disabled
 
 \else
 \makeatother

\definecolor{exampletop}{gray}{0.9}
\definecolor{examplebottom}{gray}{0.8}

\def\formtmp#1#2{{\vskip12pt\noindent\fboxsep=0pt\colorbox{#1}{\vbox{\vskip3pt\hbox to \linewidth{\hskip3pt\vbox{\raggedright\noindent\textbf{Example}~#2\vphantom{Qy}}\hfill}\vspace*{3pt}}}\par\vskip2pt%
\noindent\kern0pt}}

\newenvironment{exampleblock}[1]{\ignorespaces\def\stmtopen##1{##1}%
\formtmp{exampletop}{#1}}{\par\noindent\textcolor{examplebottom}{\rule{\columnwidth}{1pt}}\vskip2pt\par\addvspace{\baselineskip}}%

\setcopyright{acmlicensed}
\copyrightyear{2026}
\acmYear{2026}
\acmDOI{XXXXXXX.XXXXXXX}
\acmConference[MODELS '26]{ACM/IEEE International Conference on Model-Driven Engineering Languages and Systems}{October 4--9, 2026}{Malaga, Spain}
\acmISBN{978-1-4503-XXXX-X/2018/06}

\RequirePackage[
  datamodel=acmdatamodel,
  style=acmnumeric,
  ]{biblatex}

\begin{document}

\placetextbox{0.5}{0.99}{\large\colorbox{gray!3}{\textcolor{WildStrawberry}{\textbf{Author pre-print.}}}}%

\placetextbox{0.5}{0.97}{\large\colorbox{gray!3}{\textcolor{WildStrawberry}{Publication accepted for the \hreff{https://conf.researchr.org/home/models-2026}{ACM/IEEE International Conference on Model Driven Engineering Languages and Systems (MODELS'26)}.}}}%

\placetextbox{0.5}{0.05}{\colorbox{gray!3}{\textcolor{WildStrawberry}{Author pre-print. Publication accepted for} \hreff{https://conf.researchr.org/home/models-2026}{MODELS'26}.}}%

\title{Model-Driven Data Contracts for Digital Twin Services}

\author{Philipp Zech\orcidlink{0000-0002-4952-4337}}
\email{philipp.zech@uibk.ac.at}
\affiliation{%
  \institution{Department of Computer Science, University of Innsbruck}
  \city{Innsbruck}
  \country{Austria}
}
\author{Istvan David\orcidlink{0000-0002-4870-8433}}
\email{istvan.david@mcmaster.ca}
\affiliation{%
  \institution{McMaster Centre for Software Certification, McMaster University}
  \city{Hamilton}
  \country{Canada}
}
\renewcommand{\shortauthors}{Philipp Zech and Istvan David}

\begin{abstract}
Digital Twins (DT) integrate data from multiple sources. Models consume data and enable DT services such as simulations, what-if analyses, and ML-driven predictions. To ensure proper DT operation, data-driven services require data to exhibit traits such as reliability and high quality (including, e.g., accuracy, completeness, and timeliness). Yet, there is no systematic way to specify data requirements at the model level, and subsequently enact those specifications at runtime. To address this shortcoming, we propose an approach to contract-based quality management in DTs. We formally define a theory of such contracts, situate them architectually within DTs, and propose a domain-specific language to specify contracts. Our approach enables continuous data quality monitoring, thereby improving the reliability and quality of DT services.
\end{abstract}



\maketitle


\section{Introduction}\label{sec:introduction}

Digital Twins (DTs) are data-intensive systems that combine models with data from multiple sources to mirror and predict the behavior of physical systems~\cite{jones2020characterising,van2022archetypes} through services such as simulation, what-if analysis, and AI-driven prediction~\cite{miglietta2025role,rathore2021role}.
Such analytical and predictive services depend on reliable, high-quality data that further exhibits key properties, such as accuracy, completeness, and timeliness~\cite{miglietta2025role}. Models and simulations are built on assumptions about the data they consume~\cite{arthur2025simulation,paladugu2024use,tao2022digital}.

Yet, due to the lack of supporting mechanisms, languages, and tools, these data quality assumptions are not expressed in an explicit, structured, and formal way that allows for verifying or enforcing data quality~\cite{10.1007/978-3-031-70245-7_1}. As a result, these assumptions cannot be checked systematically during integration or enforced at runtime. This is not only a technical omission, but also an engineering coordination problem. In DT projects, the stakeholders who understand the assumptions made by a model are often not the stakeholders who produce, transform, and deliver the data. Typically, modeling\&simulation experts know the input properties under which model outputs remain valid; while sensor, IoT, edge, and data-platform engineers control the sources, sampling rates, preprocessing steps, and the delivery infrastructure. Therefore, data assumptions are scattered across models, code, and informal documentation, with no shared artifact that states which service requires which data property, who is expected to provide it, and how violations should be detected.
Consequently, a DT service may continue to execute on stale, incomplete, or misaligned data, producing outputs that are no longer meaningful for decision-making and compromising the validity of the DT~\cite{Gupta2025,Boschert2016}, particularly when responsibility for data and models is split across roles~\cite{AGRAWAL2023104749}.

This gap motivates \textbf{explicit data contracts for DTs} that declare the data properties required by models and the guarantees expected from data providers---all that at the boundaries of DT services. \textbf{Such contracts must be explicitly modeled}: they have to capture service-level assumptions and guarantees independently of particular sensor, middleware, or deployment technologies, while remaining precise enough to support design-time compatibility checks and runtime monitoring~\cite{kuhn2012abstracting,schmidt2006model}.

In this work, we propose the concept of data contracts for DT services, along with an initial formal framework, model-driven support, and architectural underpinnings. A data contract, in our view, is a lightweight specification of data properties required by a service to ensure its integrity, and the intended quality and validity of the results it produces. By formalizing assumptions about data quality, data contracts render these assumptions and their evaluation amenable to automation. Most importantly, data contracts enable verifying compatibility of components during integration, detecting violations during run-time, and triggering fallback or compensation mechanisms when data quality degrades.

\subsubsection*{Contributions}
In this paper, we make the following contributions.
\begin{itemize}[topsep=0pt,leftmargin=1.05em]
    \item We define the \textbf{conceptual and formal foundations} of data contracts for DT services.
    \item We define an initial \textbf{DSL for data contracts} that lets consumers specify data requirements and sources declare data guarantees.
    \item We provide the \textbf{architectural foundations} of the approach by situating it in notable DT reference architectures.
    \item We outline a \textbf{research and development map}.
\end{itemize}

The benefits of our work are twofold: (i) it provides a DSL-based engineering tool that can neatly integrate into existing engineering processes~\cite{rennels2023domain}, and (ii) it improves DT reliability through the explicit specification and automated checking of data requirements.

\subsubsection*{Running example: Smart Building DT}

We use a smart building DT as a 
running example~\cite{zech:edtconf:2025}.
This DT integrates multiple data sources and models to support building operation, e.g., simulation models to predict energy consumption from occupancy and weather data. Two common data challenges here (i) incomplete occupancy data, and (ii) delayed weather data. Both impact simulation accuracy and illustrate how data accuracy, consistency, and timeliness directly affect DT performance and decision quality.


\section{Background and Related Work}\label{sec:background}

Here, we position our work w.r.t. data quality (\secref{sec:dataquality}) and software contracts (\secref{sec:cbd}); and review the related work (\secref{sec:relwork}).

\subsection{Data quality}\label{sec:dataquality}

Data quality provides the lens through which our work views the correctness and usefulness of DT services. Rather than treating quality as an intrinsic property of a dataset, the data quality literature defines it from the perspective of consumers and intended use. Seminal works~\cite{WangStrong1996,pipino2002data,batini2009methodologies} show that data consumers evaluate quality across multiple dimensions beyond accuracy, e.g., completeness, consistency, timeliness, and currency. This consumer-centered view is particularly important for DTs: the same observation may be sufficient for one service but invalid for another. For example, coarse occupancy counts may support space-utilization analysis, but may fail detailed energy-consumption simulation. 

DT services exacerbate data quality problems by integrating heterogeneous data sources, physical-system telemetry, simulations, and AI components~\cite{mihai2022dts}. \textcite{Ouedraogo2025dtdata} identify
accuracy, completeness, and consistency as central concerns because deficiencies in these dimensions undermine the trustworthiness of DT outputs.
Therefore, data quality is context-dependent rather than absolute: its adequacy depends on the consuming service, the supported decision, and the DT's autonomy strategy~\cite{david2024dtinfonomics}. For example, a temperature reading may suffice for occupancy-level energy modeling but not for predictive maintenance.
Assessing quality thus requires relating measurable data properties to the needs of the consuming model and service

Existing data quality management techniques address low-level problems through cleansing, interpolation, imputation, etc.~\cite{Ouedraogo2025dtdata,Zahmatkesh2026imputation,batini2009methodologies}. However, these techniques do not determine quality properties required by DT services. Such a decision depends on the consumer service, the model behind it, and the characteristics of the service. Consequently, a data quality requirement in a DT must tie to the consuming service and be made explicit at the boundary where data is consumed. This motivates treating data quality properties as measurable, service-specific contract clauses in our approach.

\subsection{Contracts in Software Engineering}\label{sec:cbd}
Contracts in software engineering provide a mechanism for making boundary assumptions explicit. Design by Contract (DbC)~\cite{meyer1992applying} defines software interactions in terms of obligations and benefits: clients must satisfy preconditions, while suppliers must establish postconditions and preserve invariants. Component contract work generalizes this view beyond individual operations by distinguishing syntactic, behavioral, synchronization, and quality-of-service facets of component interaction \cite{beugnard1999making}. In both cases, the aim is an explicit specification of responsibilities at an interface, making violations attributable to the party that failed to meet its obligation.

For heterogeneous systems, assume-guarantee contracts are especially relevant. A contract pairs assumptions about the environment with guarantees provided by the component when assumptions hold. Contract theories define satisfaction, compatibility, refinement, and composition mechanisms to enable modular reasoning in which components are integrated or substituted without having to analyze the entire system~\cite{Sha20}. In these theories, a contract algebra defines operators for composing contracts across components~\cite{Ben18}. This maps naturally to DTs, where the responsibility of data production and model consumption is often split among sensor, IoT, data, simulation, and platform engineers. The consumer side needs to state assumptions about incoming data, while the producer side needs to state guarantees about outgoing data.

Contracts also connect the design-time and run-time phases of engineering processes. At design time, contract compatibility can be checked before deployment, for example by verifying that the guarantees of a source imply the assumptions of a consuming service. At runtime, monitors can observe traces or streams and detect violations of specified properties~\cite{leucker2009brief}.
Contract-based runtime verification has been applied to cyber-physical to monitor whether implementations satisfy formally specified assumptions and guarantees during operation~\cite{Nuz15}. 
Our work transfers this idea to the data boundary of DT services. A data contract specifies assumptions over input data properties and guarantees over output properties; when assumptions hold, the DT service is expected to deliver outputs with the promised qualities. Thus, data contracts complement existing contract theories by giving data-quality properties first-class status in the contract, and they motivate a model-driven notation from which design-time checks and runtime monitors can be generated.

\subsection{Related work}\label{sec:relwork}

Closest to our work are approaches connecting contract models to executable monitors, such as timing contracts for IEC 61499~\cite{Tra20b} and formal assume-guarantee contracts in CPS during operation~\cite{Nuz15}. While these techniques address timing and behavioral properties of components and controllers, they do not address data concerns in DT services as first-class assume-guarantee structures.

Other work formalizes DT artifacts using the Asset Administration Shell, OPC UA, RDF, and SHACL for machine-checkable structural constraints~\cite{Bad19,Sch19,Bar23}. These works show that DT artifacts can be checked against explicit constraints, but do not integrate behavioral assumptions and guarantees, nor turn such specifications into an end-to-end contract mechanism.

Finally, some work embeds formal verification and runtime monitoring into DTs directly by linking design models to operational data~\cite{Sar21}, synthesizing DT models from contracts~\cite{Spe20}, checking DTs against system-level timed automata contracts~\cite{Nae25}, and supervising live executions through stream monitoring and watchdogs~\cite{Bet24,Gun25}. The focus, however, remains on model correctness and temporal property satisfaction rather than data contracts for DT services.


\section{Model-driven Data Contracts for DTs}
\label{sec:contracts}

There are several reasons that motivate a model-driven approach for data contracts in DTs.
First, DTs' use for informed decision-making typically in critical domains, requires proper formal treatment of DTs' quality attributes, and the traceability, analysis, and formal checks of contracts to ensure quality data for inference.
Second, DTs are complex systems and every aspect that can be formally addressed yields a major benefit~\cite{lehner2025model}. Indeed, DT engineering exhibits some recurring patterns, e.g., how models and services exchange and consume data, that are worth capturing.
Third, DTs are similar in how we engineer them, but differ in how we use them. What stays constant across the engineering and usage phases are the models and data. This aligns with the models-and-data perspective on MDE for data-centric systems~\cite{combemale2021hitchhikers}; thus, it makes sense to model these aspects explicitly, as data contracts connect models with the data on which their services depend.
Finally, target platforms for DTs are inherently heterogeneous and---as the DT itself---evolve, thus requiring evolvability and portability of contracts~\cite{david2023towards,michael2024smart}. Given its support for abstraction and complexity management, automation through model transformations, separation of concerns, analysis, and its technology-agnostic nature, MDE is a natural enabler for our contract framework~\cite{mihai2022dts,whittle2013state}. Our choice is reinforced by the wide adoption of MDE in engineering DTs~\cite{zech2026model}.

Our proposal is uses an application-, domain-, and platform-agnostic DSL to specift data contracts for DT services. A DT service is a \emph{software component} that exposes DT functionality to external applications while hiding the DT's internal details. Contracts make the service's assumptions and guarantees explicit without requiring knowledge of its implementation.
The DSL keeps these specifications independent of the underlying technology and supports generating contract implementations for different target platforms~\cite{wkasowski2023domain}. Recent advances in LLM-based code generation render the technical implementation of contracts from models a trivial task~\cite{jiang2026survey}. 

\subsection{Formal Framework for Data Contracts}\label{sec:formal}

To formalize data contracts in a DT concept, we first need some definitions of a DT and its services.

\begin{definition}[Digital Twin]
    \label{def:dt}
    For the purposes of this paper, a Digital Twin $T = (\mathcal{M}_T, \mathcal{S}_T)$ is an abstract entity comprising a collection of models $\mathcal{M}_T = \{m_1, \dots, m_i\}$ and a collection of services $\mathcal{S}_T = \{s_1, \dots, s_j\}$.
\end{definition}

\begin{definition}[DT Service]
    \label{def:service}
    For digital twin $T=(\mathcal{M}_T,\mathcal{S}_T)$, service $s \in \mathcal{S}_T$ exposes functionality based on a collection of models $\mathcal{M}_s \subseteq \mathcal{M}_T$ through an interface. We define $s=(\mathcal{M}_s,\Sigma_s),$ where $\Sigma_s=(\mathcal{P}^s_{in}, \mathcal{P}^s_{out})$ is the service signature, $\mathcal{P}^s_{in}$ is a set of input parameters and $\mathcal{P}^s_{out}$ is a set of output parameters.
\end{definition}

The context in which a model is used is captured by the service---the same model may be exposed through different services, each with a different interface and a different data contract.

Parameters define the input and output interface of a service. Let $Name$ be a set of parameter names and $Type$ a set of data types.
\begin{definition}[Parameter]
    For service $s$, a parameter $p \in \mathcal{P}^s_{in} \cup \mathcal{P}^s_{out}$ is a typed interface declaration $p=(n,\tau) \in Name \times Type,$ where $n$ is the parameter name and $\tau$ is its type. We use the notation $n : \tau$ as a shorthand for $p=(n,\tau)$.
\end{definition}

\begin{exampleblock}{Smart building DT}%
Consider the smart building ($\mathit{sb}$) DT from the running example. We model it as
$T_{\mathit{sb}}=(\mathcal{M}_{T_{\mathit{sb}}},\mathcal{S}_{T_{\mathit{sb}}}),$
where
$\mathcal{M}_{T_{\mathit{sb}}} = \{m_e, m_f, ...\}$ contains, among others, an energy simulation model $m_e$ and a fault detection model $m_f$; and 
$\mathcal{S}_{T_{\mathit{sb}}} = \{s_e, s_f, ...\}$ contains, among others, an energy simulation service $s_e$ and a predictive maintenance service $s_f$.

Then, the energy simulation service is given by $s_e=(\{m_e\},\Sigma_e),$ where
$\Sigma_e=(\mathcal{P}^e_{in},\mathcal{P}^e_{out}),$
with
\[
\mathcal{P}^e_{in} =
o:\mathsf{OccupancyReading},
w:\mathsf{WeatherData}
\]
and
\[
\mathcal{P}^e_{out} = p:\mathsf{EnergyForecast}.
\]
Here, $o$ denotes the input parameter for occupancy data, $w$ denotes the input parameter for weather data, and $p$ denotes the output parameter for the energy prediction.
\end{exampleblock}

\begin{definition}[Service Behavior]
For a DT service $s$, let $\mathcal{B}_s$ denote the set of possible observable behaviors of $s$. A behavior $b \in \mathcal{B}_s$ represents one execution of the service, including the input data, output data, timestamps, and any metadata needed to evaluate data-quality conditions.

An implementation of a service $s$ is represented semantically by a set $\mathcal{I}_s \subseteq \mathcal{B}_s$ of behaviors that the implementation may exhibit.
\end{definition}

\begin{definition}[Data-Quality Assertion]
A data-quality assertion over a service $s$ is a pair
$\alpha=(\mathcal{P}_{\alpha},\chi_{\alpha}),$
where $\mathcal{P}_{\alpha} \subseteq \mathcal{P}^{s}_{in} \cup \mathcal{P}^{s}_{out}$ is the set of parameters targeted by the assertion, and $\chi_{\alpha}:\mathcal{B}_s \rightarrow \{\mathsf{true},\mathsf{false}\}$
is a computable predicate over service behaviors.
\end{definition}

For a finite set $X$ of data-quality assertions over $s$, its denotation is the set of behaviors satisfying all assertions $\alpha \in X$:
\[
\llbracket X \rrbracket_s
=
\{\, b \in \mathcal{B}_s
\mid
\forall \alpha \in X:\; \chi_\alpha(b)=\mathsf{true}
\,\}.
\]
By convention, if $X=\emptyset$, then $\llbracket X \rrbracket_s=\mathcal{B}_s$.

Predicate $\chi_{\alpha}$ may be implemented by any suitable computation over the observed data and metadata, such as threshold checks, windowed completeness computations, moving averages, synchronization checks, statistical tests, or comparisons against a baseline.

\begin{definition}[Data Contract]
    \label{def:contract}
    A data contract $c$ for a DT service $s$ is a tuple $c = (s, \mathcal{A}, \mathcal{G})$, where $\mathcal{A}$ is a set of data quality assertions over input parameters $\mathcal{P}^s_{in}$, and $\mathcal{G}$ is a set of data quality assertions over output parameters $\mathcal{P}^s_{out}$.

    
    We call $\mathcal{A}_c = \llbracket A \rrbracket_s$ and $\mathcal{G}_c = \llbracket G \rrbracket_s$ the assumption semantics and guarantee semantics of $c$, respectively.
\end{definition}

\begin{exampleblock}{Formal contract for the energy simulation service}%
Consider the energy simulation service \(s_e\) of the DT. Its data contract is $c_e=(s_e,\mathcal{A}_e,\mathcal{G}_e),$ where $\mathcal{A}_e$ is a set of input-side data-quality assertions and $\mathcal{G}_e$ is a set of output-side data-quality assertions.

To instantiate the data-quality assertions, we define observer functions over service behaviors. These functions extract measurable values from a behavior without changing that behavior:
\[
\mathsf{age}_o,\mathsf{age}_w : \mathcal{B}_{s_e} \rightarrow \mathbb{R}_{\geq 0},
\]
\[
\mathsf{complete}_o : \mathcal{B}_{s_e} \times \mathbb{R}_{>0} \rightarrow [0,1],
\]
and
\[
\mathsf{acc}_p : \mathcal{B}_{s_e} \rightarrow [0,1].
\]

Functions \(\mathsf{age}_o\) and \(\mathsf{age}_w\) return the age of the most recent occupancy and weather readings. Function \(\mathsf{complete}_o(b,\Delta)\) returns the completeness ratio of occupancy data in behavior \(b\) over time window \(\Delta\). Function \(\mathsf{acc}_p\) returns the accuracy of the energy forecast.

We now define the assumption-type assertions ($\mathcal{A}_e$). The first assertion requires occupancy data to be fresh:
\[
\alpha_{o,\mathit{fresh}}
=
\left(
\{o\},
\chi_{o,\mathit{fresh}}
\right),
\]
\text{where}
\[
\chi_{o,\mathit{fresh}}(b)
\Longleftrightarrow
\mathsf{age}_o(b) \leq 5\,\mathrm{min}
\]
is the computable predicate that evaluates to $\mathsf{true}$ iff the age of the occupancy reading is at most 5 minutes. Along the same lines, the other two assumptions are as follows:
\[
\alpha_{w,\mathit{fresh}}
=
\left(
\{w\},
\chi_{w,\mathit{fresh}}
\right),
\]
\[
\chi_{w,\mathit{fresh}}(b)
\Longleftrightarrow
\mathsf{age}_w(b) \leq 15\,\mathrm{min};
\]
and
\[
\alpha_{o,\mathit{complete}}
=
\left(
\{o\},
\chi_{o,\mathit{complete}}
\right).
\]
\[
\chi_{o,\mathit{complete}}(b)
\Longleftrightarrow
\mathsf{complete}_o(b,15\,\mathrm{min}) \geq 0.9.
\]
The assumption set, then, is
$
\mathcal{A}_e
=
\{
\alpha_{o,\mathit{fresh}},
\alpha_{o,\mathit{complete}},
\alpha_{w,\mathit{fresh}}
\}.$

\phantom{}

\noindent Guarantee assertions are defined in a similar vein:
\[
\gamma_{p,\mathit{acc}}
=
\left(
\{p\},
\chi_{p,\mathit{acc}}
\right);
\]
\[
\chi_{p,\mathit{acc}}(b)
\Longleftrightarrow
\mathsf{acc}_p(b) \geq 0.95,
\]
and the guarantee set is
$
\mathcal{G}_e
=
\{
\gamma_{p,\mathit{acc}}
\}.
$

\phantom{}

\noindent By the semantics of assertion sets, the assumption semantics is
\[
\begin{aligned}
\llbracket \mathcal{A}_e \rrbracket_{s_e} = \{ b \in \mathcal{B}_{s_e} \mid & \,\mathsf{age}_o(b) \leq 5\,\mathrm{min} \, \wedge \\
& \, \mathsf{complete}_o(b,15\,\mathrm{min}) \geq 0.9 \, \wedge \\
& \, \mathsf{age}_w(b) \leq 15\,\mathrm{min}\}, \\
\end{aligned}
\]
and the guarantee semantics is
\[
\llbracket \mathcal{G}_e \rrbracket_{s_e}
=
\left\{
b \in \mathcal{B}_{s_e}
\mid
\mathsf{acc}_p(b) \geq 0.95
\right\}.
\]

Thus, data contract $c_e=(s_e, \mathcal{A}_e, \mathcal{G}_e)$ states that whenever occupancy and weather inputs meet the required freshness and completeness assumptions, the service is expected to produce an energy prediction whose accuracy is at least \(0.95\).
\end{exampleblock}

Our formalization defines a precise vocabulary for data requirements and guarantees by separating the specification of data properties (the \emph{what}), from their implementation and monitoring (the \emph{how}). This allows \emph{both} for (i) reasoning about contract compatibility at design time,
and (ii) systematically enacting those contracts as runtime monitors that detect violations during operation.

To realize design-time and run-time checks, we need the following notion of satisfiability and satisfaction.

\begin{definition}[Satisfiability, Satisfaction, Met Assumptions]
Let $s$ be a DT service, $X$ a set of data-quality assertions over $s$, and $\llbracket X \rrbracket_s \subseteq \mathcal{B}_s$ the set of behaviors satisfying all assertions in $X$.
\begin{itemize}[leftmargin=1em]
    \item Assertion set $X$ is \textit{satisfiable} over $s$ iff there exists at least one behavior satisfying all assertions in $X$: $\llbracket X \rrbracket_s \neq \emptyset.$
    
    \item A behavior $b \in \mathcal{B}_s$ \textit{satisfies} $X$ iff $b \in \llbracket X \rrbracket_s.$ (Written as $b \models_s X$.)
    
    \item For data contract $c=(s,\mathcal{A},\mathcal{G})$, assumptions are \emph{met} by a behavior $b$ iff $b \models_s \mathcal{A},$ and guarantees are \textit{met} by $b$ iff $b \models_s \mathcal{G}.$ Equivalently, we define the runtime predicates $\mathsf{assumptionMet}_c(b) \Longleftrightarrow b \in \llbracket \mathcal{A} \rrbracket_s$ and $\mathsf{guaranteeMet}_c(b) \Longleftrightarrow b \in \llbracket \mathcal{G} \rrbracket_s.$

    \item An implementation $\mathcal{I}_s \subseteq \mathcal{B}_s$ \textit{satisfies} contract $c$, written $\mathcal{I}_s \models c$, iff every implementation behavior that meets the assumptions also meets the guarantees: $\mathcal{I}_s \models c \Longleftrightarrow \mathcal{I}_s \cap \llbracket \mathcal{A} \rrbracket_s \subseteq \llbracket \mathcal{G} \rrbracket_s.$
\end{itemize}
\end{definition}

Thus, assumption satisfiability is used at design time as a sanity check to ensure an assumption set is possible, i.e., that $\llbracket \mathcal{A} \rrbracket_s \neq \emptyset$.
At runtime, we observe one concrete behavior $b$ and we check whether the observed behavior meets the assumptions: $\mathsf{assumptionMet}_c(b) \Longleftrightarrow b \in \llbracket \mathcal{A} \rrbracket_s.$
Contract satisfaction is even stronger: it states that, for all behaviors an implementation may exhibit, whenever the assumptions are met, the guarantees are met as well, i.e., $\mathcal{I}_s \models c \Longleftrightarrow \mathcal{I}_s \cap \llbracket \mathcal{A} \rrbracket_s \subseteq \llbracket \mathcal{G} \rrbracket_s.$

In addition, we define the notion of compatibility for design-time integration checks as follows.

\begin{definition}[Compatibility]
Let $c=(s,\mathcal{A},\mathcal{G})$ be a data contract for service \(s\), and let
$\mathcal{D}_{\mathit{src}} \subseteq \mathcal{B}_s$ denote the set of behaviors guaranteed by the selected data sources, projected onto the input interface of $s$. The data sources are \emph{compatible} with $c$ iff $\mathcal{D}_{\mathit{src}} \subseteq \llbracket \mathcal{A} \rrbracket_s.$
\end{definition}

\begin{exampleblock}{Checking contracts}%
For the energy simulation contract $c_e=(s_e, \mathcal{A}_e, \mathcal{G}_e)$, a design-time consistency check asks whether the assumption set is satisfiable, i.e., 
$
\llbracket \mathcal{A}_e \rrbracket_{s_e} \neq \emptyset.
$
This holds, e.g., with behavior \(b \in \mathcal{B}_{s_e}\) such that
\[
\begin{aligned}
\,\mathsf{age}_o(b) &= 4\,\mathrm{min} \, \wedge \\
\, \mathsf{complete}_o(b,15\,\mathrm{min}) &= 0.95 \, \wedge \\
\, \mathsf{age}_w(b) &= 10\,\mathrm{min}\ \\
\end{aligned}
\]
This behavior satisfies all assumptions and therefore, it witnesses that the assumption set is satisfiable.

\phantom{}

\noindent A design-time compatibility check is relative to the data source. Suppose the occupancy and weather data providers jointly guarantee behavior set
\[
\begin{aligned}
\mathcal{D}_{\mathit{src}} = \{ b \in \mathcal{B}_{s_e} \mid & \,\mathsf{age}_o(b) \leq 3\,\mathrm{min} \, \wedge \\
& \, \mathsf{complete}_o(b,15\,\mathrm{min}) \geq 0.95 \, \wedge \\
& \, \mathsf{age}_w(b) \leq 10\,\mathrm{min}\}, \\
\end{aligned}
\]
Since
$
\mathcal{D}_{\mathit{src}} \subseteq \llbracket \mathcal{A}_e \rrbracket_{s_e},
$
the selected data sources are compatible with the energy simulation contract.
\end{exampleblock}

\tabref{tab:checks} summarizes how these formal concepts map onto various checks at various lifecycle phases of a DT service.

\begin{table*}[]
\setlength{\tabcolsep}{2pt}
\caption{Contract validation and verification activities (DT = design-time, RT = run-time)}
\label{tab:checks}
{
\footnotesize
\begin{tabular}{@{}llll@{}}
\toprule
\textbf{Concept} & \textbf{Phase} & \textbf{Formal condition} & \textbf{Question answered} \\
\midrule

Assertion-set satisfiability & DT & $\llbracket X \rrbracket_s \neq \emptyset$ & Is the assertion set $X$ internally possible? \\

Assumption-set satisfiability & DT & $\llbracket \mathcal{A} \rrbracket_s \neq \emptyset$ & Are the contract assumptions internally possible? \\

Source compatibility & DT & $\mathcal{D}_{\mathit{src}} \subseteq \llbracket \mathcal{A} \rrbracket_s$ & Do the selected data sources guarantee data that satisfies the service assumptions? \\

Assumptions met & RT & $\mathsf{assumptionMet}_c(b) \Longleftrightarrow b \in \llbracket \mathcal{A} \rrbracket_s$ & Does the currently observed behavior satisfy the input-side assumptions? \\

Guarantees met & RT & $\mathsf{guaranteeMet}_c(b) \Longleftrightarrow b \in \llbracket \mathcal{G} \rrbracket_s$ & Does the currently observed behavior satisfy the output-side guarantees? \\

Contract satisfaction & DT/RT & $\mathcal{I}_s \models c \Longleftrightarrow \mathcal{I}_s \cap \llbracket \mathcal{A} \rrbracket_s \subseteq \llbracket \mathcal{G} \rrbracket_s$ & Does the implementation guarantee the promised output quality whenever its assumptions are met? \\

\bottomrule
\end{tabular}
}
\end{table*}

\subsection{A DSL for Data Contracts}\label{ssec:dsl}

\begin{figure}[bth]
    \centering
    \includegraphics[width=\columnwidth]{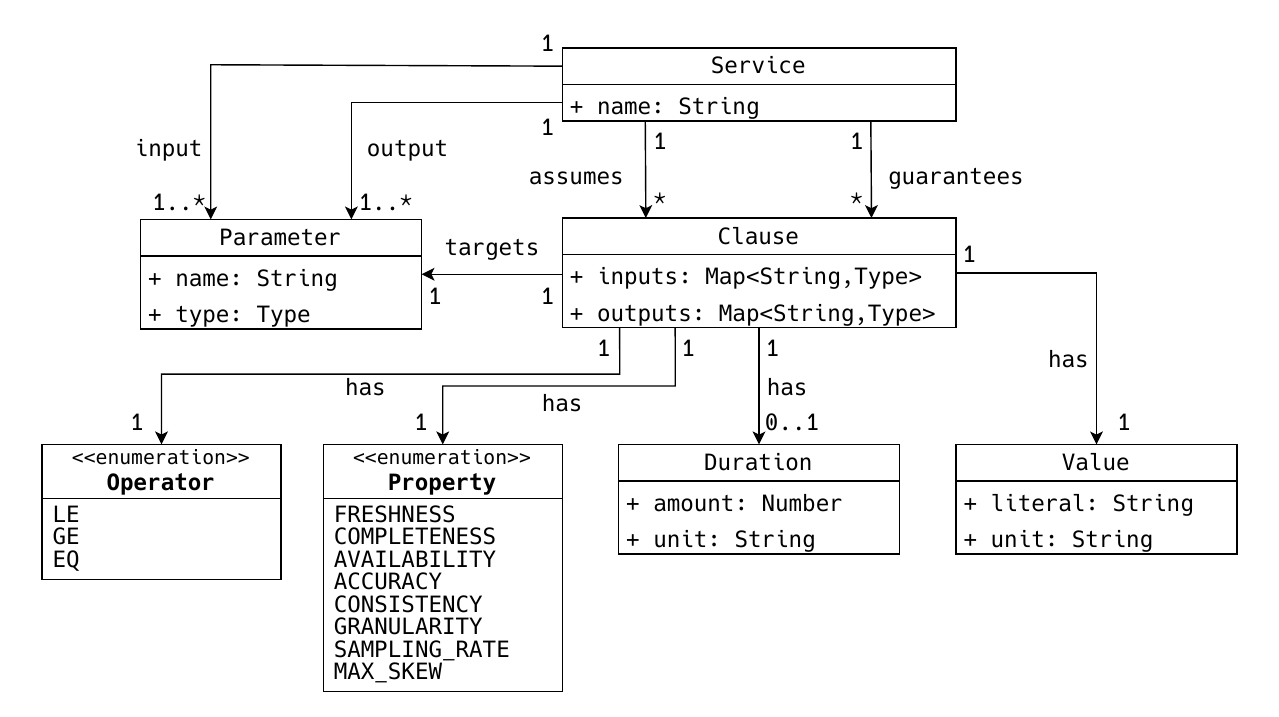}
    \caption{Metamodel of data contracts for DT services.}
    \label{fig:dsl}
\end{figure}

The formal definitions in \secref{sec:formal} map directly onto a DSL (cf.~\figref{fig:dsl}). A \emph{contract} is expressed as a named \texttt{Service} with typed I/O parameters. Each parameter can be targeted by one or more \texttt{Clause}s, which are grouped into \texttt{assumes} and \texttt{guarantees} aggregations. A \texttt{Clause} binds a \texttt{Property} (e.g., \textsc{Freshness}, \textsc{Completeness}, \textsc{Accuracy}) to a relational \texttt{Operator} and a typed \texttt{Value}, optionally scoped to a \texttt{Duration} window. Our platform-agnostic DSL keeps the specification decoupled from any particular DT infrastructure and enables generation of actual contract implementations~\cite{wkasowski2023domain}.

\subsubsection{Example: Contract expressed in the DSL (energy simulation service)}

\listref{lst:energy} shows the contract for the energy simulation service (corresponding to Example 1), expressed in the DSL.

\begin{lstlisting}[
  language=ContractDSL,
  numbers=left,
  caption={Contract in DSL (energy simulation service)},
  label={lst:energy}]
service EnergySimulation {
  inputs:  occupancy : OccupancyReading
           weather   : WeatherData
  outputs: prediction: EnergyForecast

  assumes {
    occupancy : FRESHNESS    <= 5min
                COMPLETENESS >= 0.9 over 15min
    weather   : FRESHNESS    <= 15min
  }
  guarantees {
    prediction: ACCURACY >= 0.95
  }
}
\end{lstlisting}

\emph{At design time}, our contracts can be used for model-checking to verify that the guarantees of data sources satisfy the assumptions of services and thus, identify incompatibilities before deployment. \emph{At runtime}, monitors generated from these contracts enforce them over live data streams and raise a violation if a constraint is breached. This turns what would otherwise be silent degradation into an \emph{explicit and traceable event} that the DT can act on.

\subsection{Architectural underpinnings}\label{ssec:architecture}

Data contracts span the boundary between data producers and data consumers. \textbf{ISO/IEC 30173}
defines a DT as a digital representation of a real-world entity that is connected to that entity and provides services based on its data. This framing makes the data boundary explicit. The connection between the physical entity and its digital representation is precisely where data quality properties must be specified and enforced. We situate this boundary in two reference architectures widely used in DT and Industry 4.0 contexts.

In \textbf{ISO 23247},
the reference architecture defines a Device Communication Entity (DCE) and a Digital Twin Entity (DTE). DT services reside in the DTE, consuming data forwarded by the DCE and exposing results to the User Entity (UE). A data contract is attached to a DT service: \emph{assumptions} specify the data quality properties the service requires on its inputs, and \emph{guarantees} specify the properties the service promises on its outputs to the UE and other services. Contract checking at design time verifies compatibility at both interfaces, and runtime monitors enforce them during operation.

In \textbf{RAMI 4.0},
DT services reside in the Functional layer. A data contract is, again, attached to the service: \emph{assumptions} specify what quality properties the service requires from data flowing from the Communication and Information layers; \emph{guarantees} specify what the service promises to components above. Violations detected at runtime manifest in the Functional layer and trigger compensations.

\section{Discussion}\label{sec:discussion}

The smart building DT from \secref{sec:introduction} illustrates how data contracts apply across the full life cycle of a DT. Consider first the deployment stage. An energy optimization model requires occupancy, HVAC
state, and weather data at defined rates and granularity. These assumptions are typically implicit, buried in model code or documentation. A contract makes them explicit at the service interface. Before deployment, a compatibility check at the DT boundary verifies that available data sources satisfy those assumptions and catches misconfigurations before it causes wrong outputs.

The same logic applies during system integration when the DT integrates subsystems from different vendors with inconsistent formats and update semantics. Each service declares what it requires and guarantees, and compatibility is checked at the communication layers (cf.~\secref{ssec:architecture}). Mismatches surface during composition rather than at runtime, where they are far more costly to fix.

Once operational, contracts shift from design-time artifacts to runtime monitors. A fault detection model that receives stale or incomplete telemetry is a typical case. A monitor generated from a contract checks each incoming stream against the required data properties. A violation raises an explicit, traceable event rather than letting the model fail silently. That traceability also enables controlled failover. When a data feed fails, the contract identifies exactly which properties are violated and provides a formal basis for comparing alternative sources. The DT can switch to a substitute or fail gracefully, rather than continuing with corrupted outputs.

Contracts also remain useful as the DT evolves. For example, when a building subsystem is upgraded and its data interface changes, all downstream dependencies are explicit in the contracts of affected services. Compatibility can be re-checked before the updated subsystem goes into operation, making upgrade impact traceable and reducing regression risk, a property that becomes increasingly important as DTs grow in scope~\cite{margaria2009continuous}.

These scenarios share a common theme: the absence of explicit data contracts forces engineers to rely on implicit assumptions, manual checks, or hard-coded logic. Alas, contracts do not eliminate underlying complexity. Sensors still fail, interfaces still change, and data continues to degrade. However, contracts still help by establishing the conditions for correct operation in a visible, checkable, and actionable way. This shift from implicit to explicit and from reactive to proactive is what our proposal fundamentally offers.

\subsection{Benefits}

\noindent\textbf{Explicit data dependencies.} Contracts make the assumptions between data producers and model consumers machine-readable. This is a prerequisite for any systematic quality management and replaces implicit, undocumented expectations~\cite{giatzis2024software}.

\noindent\textbf{Design-time compatibility checking.} Before deployment, contracts can be used for model checking to verify that data source guarantees satisfy service assumptions. Incompatibilities are caught early, when they are cheapest to fix~\cite{schmidt2006model}.

\noindent\textbf{Runtime violation detection.} Monitors generated directly from contracts enforce quality thresholds over live data streams. Violations trigger defined responses rather than silent degradation, improving DT reliability and trustworthiness~\cite{Bencomo2014,vierhauser2021towards}.

\noindent\textbf{Code generation.} The DSL decouples contract specification from implementation. Contract monitors can be generated for specific target platforms~\cite{burgueno2025automation}, keeping specifications reusable across different DTs and removing manual instrumentation effort~\cite{whittle2013state}.

\noindent\textbf{Model-driven integration and discovery.} Federated DTs often span heterogeneous technical and organizational boundaries~\cite{david2024interoperability}. Contracts merit their integration by automated matching of producers to consumers, discovery of back-up producers upon failure~\cite{adesanya2026systems}, and generation of producers stubs for continued operation~\cite{zhu2023stubcoder}.

\subsection{Challenges}

\noindent\textbf{DSL coverage and expressiveness.} The current DSL covers a core set of data quality properties which are capture as a fixed enumeration. This design choice prioritizes simplicity and strong typing for common use cases but inherently limits extensibility, e.g., adding a new, domain-specific property (e.g., \texttt{SensorCalibrationStatus}) requries touching the grammar. A key challenge for future work thus is to evolve our DSL into a more open model, e.g., by replacing the enumeration with a registry-based approach, where new properties can be defined and registered along with their validation logic (e.g., lambda functions). In addition, extending the DSL to composite and temporal constraints also requires further design work and empirical validation with practitioners. Getting the language right is central. DSLs are only effective when they are expressive enough to capture real needs yet simple enough to be adopted~\cite{wkasowski2023domain}.

\noindent\textbf{Discovery of congruent data sources.} Matching data sources to service assumptions at scale requires discovery mechanisms that go beyond static configuration. Model-driven techniques for automated source-to-consumer matching employing semantic descriptions and ontologies~\cite{combemaleo2025challenges} are a natural direction, and the technology-agnostic nature of the DSL makes such integration feasible.

\noindent\textbf{Coordination and compensation mechanisms.} When a violation is detected, the DT must respond accordingly. Specifying and generating coordination logic, e.g., using state machines or model transformations, to trigger fallback modes, switch sources, or degrade gracefully is a key open challenge. This is precisely where MDE has the most to offer by making coordination logic a first-class, generatable artifact rather than handwritten glue code~\cite{nieves2011coordination}.

\noindent\textbf{Scalability and runtime overhead.} Continuous monitoring of data quality constraints over voluminous streams introduces overhead. Efficient monitor synthesis and selective enforcement strategies~\cite{casolari2010selection,kolovos2009grand} need to be evaluated at scale in realistic DT settings.

\noindent\textbf{Tooling and integration.} The approach requires tooling for contract authoring, compatibility checking, monitor generation, and violation handling. Building and validating this model-driven toolchain in realistic DT settings is the most immediate next step. Model-based tool integration provides a feasible path forward~\cite{kapsammer2006model}.

\noindent\textbf{Observability for contract enforcement} Our approach hinges on the observability of a DT. To enforce properties like \texttt{FRESHNESS} or \texttt{SAMPLING\_RATE}, runtime monitors require access to, e.g., timestamps and source identifiers, along the raw data. This implies a foundational requirement for any \textit{contract-ready} DT platform. We recommend that emerging DT standards define a minimal set of observability capabilities to ensure contract-readiness of DTs.


\section{Future Plans}
\label{sec:conclusion}

Our current goal is to implement a working prototype covering contract authoring, compatibility checking, and runtime monitor generation from DSL specifications. Following this, we plan to (i) validate the toolchain in realistic DT settings through case studies with practitioners; (ii) modify and extend the DSL to enable on-the-fly integration of additional domain-specific properties, composite constraints, and temporal dependencies; (iii) develop automated source-to-consumer matching to discover compatible data sources at integration and switch between them at runtime; (iv) generate coordination and compensation logic directly from contract violations using state machines or model transformations, making graceful degradation a designable, generatable behavior; (v) evaluate the approach quantitatively, assessing the impact of contract enforcement on DT reliability and decision quality. These steps form a concrete path toward data quality as a first-class, model-driven engineering concern in DTs; and (vi) define a formal algebra for data contracts to enable compositional reasoning to derive end-to-end data quality guarantees from individual contracts at the service level.


\begin{acks}
This research has been funded by the Austrian Research Promotion Agency (FFG), 927874; and the Natural Sciences and Engineering Research Council of Canada (NSERC), DGECR-2024-00293.
\end{acks}


\printbibliography

\end{document}

%% file: preprint-commands.tex
\usepackage{eso-pic}
\definecolor{linkblue}{HTML}{006fbd}
\newcommand{\hreffinternal}[3]{\href{#1}{\textcolor{#3}{#2}}}
\newcommand{\hreff}[2]{\hreffinternal{#1}{#2}{linkblue}}

\newcommand{\placetextbox}[3]{
  \setbox0=\hbox{#3}
  \AddToShipoutPictureFG*{
    \put(\LenToUnit{#1\paperwidth},\LenToUnit{#2\paperheight}){\vtop{{\null}\makebox[0pt][c]{#3}}}%
  }%
}%

%% file: references.bib
@book{Bencomo2014,
  title = {{Models\@run.time: Foundations, Applications, and Roadmaps}},
  author = {Bencomo, Nelly and others},
  publisher = {Springer},
  series = {LNCS},
  year = {2014},
  volume = {8378}
}

@article{van2022archetypes,
	title        = {{Archetypes of Digital Twins}},
	author       = {Van Der Valk, Hendrik and Ha{\ss}e, Hendrik and M{\"o}ller, Frederik and Otto, Boris},
	year         = 2022,
	journal      = {Business \& Inf Sys Eng},
	publisher    = {Springer},
	volume       = 64,
	number       = 3,
	pages        = {375--391},
}

@article{jones2020characterising,
	title        = {{Characterising the Digital Twin: A systematic literature review}},
	author       = {Jones, David and others},
	year         = 2020,
	journal      = {CIRP J Manuf Sci Technol},
	publisher    = {Elsevier},
	volume       = 29,
	pages        = {36--52},
}

@article{rathore2021role,
	title        = {{The Role of AI, Machine Learning, and Big Data in Digital Twinning: A Systematic Literature Review, Challenges, and Opportunities}},
	author       = {Rathore, M Mazhar and Shah, Syed Attique and Shukla, Dhirendra and Bentafat, Elmahdi and Bakiras, Spiridon},
	year         = 2021,
	journal      = {IEEE Access},
	publisher    = {IEEE},
	volume       = 9,
	pages        = {32030--32052},
}

@inproceedings{miglietta2025role,
	title        = {{The Role of Data in Digital Twins: Value Creation and Interoperability}},
	author       = {Miglietta, Mariasimona and others},
	year         = 2025,
	booktitle    = {Intl. Conf. on Ext. Reality},
	pages        = {235--254},
	organization = {Springer}
}

@inproceedings{paladugu2024use,
	title        = {{The Use of Computational Modeling and Simulation to Design and Evaluate a Distributed Work System}},
	author       = {Paladugu, A and Fernandes, A and IJtsma, M},
	year         = 2024,
	booktitle    = {Proc Hum Factors Ergon Soc Annu Meet},
	volume       = 68,
	number       = 1,
	pages        = {243--249},
	organization = {SAGE}
}

@article{arthur2025simulation,
	title        = {{Simulation Project Quality and Validation Profiling}},
	author       = {Arthur, Daniel JW and Winch, Graham W},
	year         = 2025,
	journal      = {Journal of Simulation},
	publisher    = {Taylor \& Francis},
	volume       = 19,
	number       = 6,
	pages        = {694--710},
}

@article{tao2022digital,
	title        = {{Digital Twin Modeling}},
	author       = {Tao, Fei and Xiao, Bin and Qi, Qinglin and Cheng, Jiangfeng and Ji, Ping},
	year         = 2022,
	journal      = {J Manuf Syst},
	publisher    = {Elsevier},
	volume       = 64,
	pages        = {372--389},
}

@inbook{Boschert2016,
	title        = {{Digital Twin---The Simulation Aspect}},
	author       = {Boschert, Stefan and Rosen, Roland},
	year         = 2016,
	booktitle    = {Mechatronic Futures: Challenges and Solutions for Mechatronic Systems and their Designers},
	publisher    = {Springer},
	pages        = {59--74},
}

@incollection{Gupta2025,
	title        = {{Digital Twin Challenges}},
	author       = {Gupta, Sunil and Iyer, Ravi S. and Kumar, Sanjeev},
	year         = 2025,
	booktitle    = {Digital Twins: Advancements in Theory, Implementation, and Applications},
	publisher    = {Springer},
	pages        = {19--42},
}

@inproceedings{10.1007/978-3-031-70245-7_1,
	title        = {{Requirements Engineering for Digital Twins: a Cross-Domain Systematic Literature Review}},
	author       = {Aurora Mac{\'i}as and Elena Navarro and Carlos E. Cuesta and Uwe Zdun},
	year         = 2025,
	booktitle    = {Quality of Information and Communications Technology},
	publisher    = {Springer},
}

@article{AGRAWAL2023104749,
	title        = {{Digital Twin: Where Do Humans Fit In?}},
	author       = {Ashwin Agrawal and others},
	year         = 2023,
	journal      = {Automation in Construction},
	volume       = 148,
	pages        = 104749,
}

@article{Nuz15,
	title        = {{A Platform-Based Design Methodology With Contracts and Related Tools for the Design of Cyber-Physical Systems}},
	author       = {Nuzzo, Pierluigi and Sangiovanni-Vincentelli, Alberto L. and Bresolin, Davide and Geretti, Luca and Villa, Tiziano},
	year         = 2015,
	journal      = {Proc. of the IEEE},
	volume       = 103,
	number       = 11,
	pages        = {2104--2132},
}

@misc{adesanya2026systems,
  title = {Systems of Twinned Systems: A Systematic Literature Review},
  author = {Feyi Adesanya and Kanan Castro Silva and Valdemar V. Graciano Neto and Istvan David},
  year = {2026},
  eprint = {2505.19916},
}

@article{Ben18,
	title        = {{Contracts for System Design}},
	author       = {Benveniste, Albert and Caillaud, Beno{\^\i}t and Nickovic, Dejan and Passerone, Roberto and Raclet, Jean-Baptiste and Reinkemeier, Philipp and Sangiovanni-Vincentelli, Alberto and Damm, Werner and Henzinger, Thomas A and Larsen, Kim G},
	year         = 2018,
	journal      = {Foundations and trends in electronic design automation},
	publisher    = {Emerald},
	volume       = 12,
	number       = {2-3},
	pages        = {124--400},
}

@article{Sha20,
	title        = {{Assume/guarantee contracts for dynamical systems: Theory and computational tools}},
	author       = {Sharf, Miel and Besselink, Bart and Molin, Adam and Zhao, Qiming and Johansson, Karl Henrik},
	year         = 2021,
	journal      = {IFAC-PapersOnLine},
	publisher    = {Elsevier},
	volume       = 54,
	number       = 5,
	pages        = {25--30},
}

@inproceedings{Tra20b,
	title        = {{Timing Contracts and Monitors for Safety Relevant Controller Design in IEC 61499}},
	author       = {Tran, Duc Do and Grüttner, Kim and Oppenheimer, Frank and Nebel, Wolfgang},
	year         = 2020,
	booktitle    = {2020 25th IEEE Intl. Conf. on Emerging Technologies and Factory Automation (ETFA)},
	volume       = 1,
	number       = {},
	pages        = {156--163},
}

@inproceedings{Bad19,
	title        = {{The Semantic Asset Administration Shell}},
	author       = {Bader, Sebastian R. and Maleshkova, Maria},
	year         = 2019,
	booktitle    = {Semantic Systems. The Power of AI and Knowledge Graphs},
	publisher    = {Springer},
	pages        = {159--174},
}

@inproceedings{Sch19,
	title        = {{A Formal Mapping Between OPC UA and the Semantic Web}},
	author       = {Schiekofer, Rainer and Grimm, Stephan and Brandt, Maja Milicic and Weyrich, Michael},
	year         = 2019,
	booktitle    = {2019 IEEE 17th Intl. Conf. on Industrial Informatics (INDIN)},
	volume       = 1,
	number       = {},
	pages        = {33--40},
}

@article{Bar23,
	title        = {{Deriving semantic validation rules from industrial standards: An OPC UA study}},
	author       = {Yashoda Saisree Bareedu and others},
	year         = 2024,
	journal      = {Semantic Web},
	volume       = 15,
	number       = 2,
	pages        = {517--554},
}

@inproceedings{Sar21,
	title        = {{A Digital Twin with Runtime-Verification for Industrial Development-Operation Integration}},
	author       = {Saratha, Siva D. Chandrasekaran and Grimm, Christoph and Wawrzik, Frank},
	year         = 2021,
	booktitle    = {IEEE Intl Conf on Eng, Tech and Inno},
	pages        = {1--9},
}

@inproceedings{Spe20,
	title        = {{Production Recipe Validation through Formalization and Digital Twin Generation}},
	author       = {Spellini, Stefano and Chirico, Roberta and Panato, Marco and Lora, Michele and Fummi, Franco},
	year         = 2020,
	booktitle    = {2020 Design, Automation \& Test in Europe Conference \& Exhibition (DATE)},
	volume       = {},
	number       = {},
	pages        = {1698--1703},
}

@article{Nae25,
	title        = {{Contract-Based Verification of Digital Twins}},
	author       = {Naeem, Muhammad and Seceleanu, Cristina},
	year         = 2026,
	booktitle    = {Engineering of Complex Computer Systems},
	publisher    = {Springer},
	pages        = {338--357},
}

@inproceedings{Bet24,
	title        = {{Digital Twin Enabled Runtime Verification for Autonomous Mobile Robots under Uncertainty}},
	author       = {Betzer, Joakim Schack and Boudjadar, Jalil and Frasheri, Mirgita and Talasila, Prasad},
	year         = 2024,
	booktitle    = {Intl. Symp. on Distributed Simulation and Real Time Applications},
	pages        = {10--17}
}

@article{Gun25,
  title = {Verification of Digital Twins using Classical and Statistical Model Checking},
  author = {Gunasekaran, Raghavendran and Haverkort, Boudewijn},
  year = {2025},
  journal = {Electron Proc Theor Comput Sci},
  publisher = {Open Publishing Association},
  volume = {418},
  pages = {16–23},
}

@article{Ouedraogo2025dtdata,
	title        = {{Digital Twin Data Management: A Comprehensive Review}},
	author       = {Ouedraogo, Ezekiel B. and others},
	year         = 2025,
	journal      = {IEEE Trans. on Big Data},
	volume       = 11,
	number       = 5,
	pages        = {2224--2243},
}

@inproceedings{david2023towards,
  title = {Towards a Taxonomy of Digital Twin Evolution for Technical Sustainability},
  author = {Istvan David and Dominik Bork},
  year = {2023},
  booktitle = {{ACM/IEEE} International Conference on Model Driven Engineering Languages and Systems Companion, {MODELS-C}},
  publisher = {{IEEE}},
  pages = {934--938},
}

@inproceedings{david2024dtinfonomics,
	title        = {{Infonomics of Autonomous Digital Twins}},
	author       = {David, Istvan and Bork, Dominik},
	year         = 2024,
	booktitle    = {Advanced Information Systems Engineering},
	publisher    = {Springer},
	pages        = {563--578},
}

@article{Zahmatkesh2026imputation,
	title        = {{Spatio-Temporal Missing Data Imputation: A Systematic Literature Review with a Focus on Statistical and Machine Learning-Based Approaches}},
	author       = {Zahmatkesh, Samira and Zech, Philipp},
    journal      = {ACM Comput Surv},
	year         = 2026,
	publisher    = {ACM},
	volume       = 58,
	number       = 10,
	numpages     = 41
}

@article{mihai2022dts,
	title        = {{Digital Twins: A Survey on Enabling Technologies, Challenges, Trends and Future Prospects}},
	author       = {Mihai, Stefan and Yaqoob, Mahnoor and Hung, Dang V. and Davis, William and Towakel, Praveer and Raza, Mohsin and Karamanoglu, Mehmet and Barn, Balbir and Shetve, Dattaprasad and Prasad, Raja V. and Venkataraman, Hrishikesh and Trestian, Ramona and Nguyen, Huan X.},
	year         = 2022,
	journal      = {IEEE Communications Surveys \& Tutorials},
	volume       = 24,
	number       = 4,
	pages        = {2255--2291}
}

@inproceedings{kuhn2012abstracting,
	title        = {{Lessons learned from evaluating MDE abstractions in an industry field study}},
	author       = {Kuhn, Adrian and Murphy, Gail C.},
	year         = 2012,
	booktitle    = {Intl Workshop on Experiences and Empirical Studies in Software Modelling},
	publisher    = {ACM},
	articleno    = 3,
	numpages     = 5
}

@book{wkasowski2023domain,
	title        = {{Domain-Specific Languages}},
	author       = {W{\k{a}}sowski, Andrzej and Berger, Thorsten},
	year         = 2023,
	publisher    = {Springer},
}

@article{jiang2026survey,
	title        = {{A Survey on Large Language Models for code Generation}},
	author       = {Jiang, Juyong and others},
	year         = 2026,
	journal      = {ACM Trans Softw Eng Method},
	volume       = 35,
	number       = 2,
	pages        = {1--72}
}

@inproceedings{kapsammer2006model,
  title={{Model-based Tool Integration\textemdash State of the Art and Future Perspectives}},
  author={Kapsammer, Elisabeth and Reiter, Thomas and Schwinger, Wieland},
  booktitle={Proc. of the 3rd Intl. Conf. on Cybernetics and Inf. Technologies, Systems and Applications},
  year={2006}
}

@incollection{casolari2010selection,
  title={{On the Selection of Models for Runtime Prediction of System Resources}},
  author={Casolari, Sara and Colajanni, Michele},
  booktitle={Run-time Models for Self-managing Systems and Applications},
  pages={25--44},
  year={2010},
  publisher={Springer}
}

@incollection{nieves2011coordination,
  title={{Coordination, Organisation and Model-Driven Approaches for Dynamic, Flexible, Robust Software and Services Engineering}},
  author={Nieves, Juan Carlos and others},
  booktitle={Service engineering: European research results},
  pages={85--115},
  year={2011},
  publisher={Springer}
}

@inproceedings{combemaleo2025challenges,
  title={{On the Challenges of Integrating Digital Twins}},
  author={Combemale, Benoit and others},
  booktitle={2025 ACM/IEEE 28th Intl. Conf. on Model Driven Engineering Languages and Systems Companion},
  pages={243--249},
  year={2025},
  organization={IEEE}
}

@article{combemale2021hitchhikers,
  title = {A Hitchhiker's Guide to Model-Driven Engineering for Data-Centric Systems},
  author = {Combemale, Benoit and others},
  year = {2021},
  journal = {IEEE Software},
  volume = {38},
  number = {4},
  pages = {71--84},
}

@inproceedings{kolovos2009grand,
	title        = {{The Grand Challenge of Scalability for Model Driven Engineering}},
	author       = {Kolovos, Dimitrios S. and Paige, Richard F. and Polack, Fiona A. C.},
	year         = 2009,
	booktitle    = {Models in Software Engineering},
	publisher    = {Springer},
	pages        = {48--53},
}

@inproceedings{giatzis2024software,
  title={{Software Engineering Practices in Smart Contract Development: A Systematic Mapping Study}},
  author={Giatzis, Antonios and Arvanitou, Elvira-Maria and Papadopoulou, Danai and Maikantis, Theodoros and Nikolaidis, Nikolaos and Feitosa, Daniel and Georgiadis, Christos and Ampatzoglou, Apostolos and Chatzigeorgiou, Alexander and Konstantinidis, Evdokimos and others},
  booktitle={Intl. Conf. on Product-Focused Software Process Improvement},
  pages={360--367},
  year={2024},
  organization={Springer}
}

@article{schmidt2006model,
  title={{Model-driven Engineering}},
  author={Schmidt, Douglas C and others},
  journal={Computer},
  volume={39},
  number={2},
  pages={25},
  year={2006},
  publisher={IEEE}
}

@inproceedings{vierhauser2021towards,
  title={{Towards a Model-Integrated Runtime Monitoring Infrastructure for Cyber-Physical Systems}},
  author={Vierhauser, Michael and Marah, Hussein and Garmendia, Antonio and Cleland-Huang, Jane and Wimmer, Manuel},
  booktitle={2021 IEEE/ACM 43rd Intl. Conf. on software engineering: new ideas and emerging results (ICSE-NIER)},
  pages={96--100},
  year={2021},
  organization={IEEE}
}

@article{zhu2023stubcoder,
  title={Stubcoder: Automated generation and repair of stub code for mock objects},
  author={Zhu, Hengcheng and others},
  journal={ACM Trans Softw Eng Method},
  volume={33},
  number={1},
  pages={1--31},
  year={2023},
  publisher={ACM New York, NY, USA}
}

@article{burgueno2025automation,
  title={{Automation in Model-Driven Engineering: A Look Back, and Ahead}},
  author={Burgue{\~n}o, Lola and Di Ruscio, Davide and Sahraoui, Houari and Wimmer, Manuel},
  journal={ACM Trans Softw Eng Method},
  volume={34},
  number={5},
  pages={1--25},
  year={2025},
  publisher={ACM New York, NY}
}

@article{margaria2009continuous,
  title={{Continuous Model-driven Engineering}},
  author={Margaria, Tiziana and Steffen, Bernhard},
  journal={Computer},
  volume={42},
  number={10},
  pages={106--109},
  year={2009},
  publisher={IEEE}
}

@inproceedings{michael2024smart,
  title = {Digital Twin Evolution for Sustainable Smart Ecosystems},
  author = {Michael, Judith and David, Istvan and Bork, Dominik},
  year = {2024},
  booktitle = {{ACM/IEEE} International Conference on Model Driven Engineering Languages and Systems Companion, {MODELS-C}},
  publisher = {{ACM}},
  pages = {1061–1065},
}

@InProceedings{zech:edtconf:2025,
  author    = {Zech, Philipp and others},
  booktitle = {{ACM/IEEE Intl. Conf. on Model-Driven Engineering Lang and Sys Comp.}},
  title     = {{Model-driven Digital Twins for AECO}},
  year      = {2025},
  pages     = {224-235},
  publisher = {IEEE},
}

@article{meyer1992applying,
  title = {Applying 'design by contract'},
  author = {Meyer, B.},
  year = {1992},
  journal = {Computer},
  volume = {25},
  number = {10},
  pages = {40--51},
}

@article{beugnard1999making,
  title = {Making components contract aware},
  author = {Beugnard, A. and Jezequel, J.-M. and Plouzeau, N. and Watkins, D.},
  year = {1999},
  journal = {Computer},
  volume = {32},
  number = {7},
  pages = {38--45},
}

@article{WangStrong1996,
	title        = {{Beyond Accuracy: What Data Quality Means to Data Consumers}},
	author       = {Richard Y. Wang and Diane M. Strong},
	year         = 1996,
	journal      = {J Manag Inf Syst},
	publisher    = {Taylor \& Francis},
	volume       = 12,
	number       = 4,
	pages        = {5--33},
}

@article{leucker2009brief,
  title = {A brief account of runtime verification},
  author = {Martin Leucker and Christian Schallhart},
  year = {2009},
  journal = {The Journal of Logic and Algebraic Programming},
  volume = {78},
  number = {5},
  pages = {293--303},
}

@article{pipino2002data,
  title = {Data quality assessment},
  author = {Pipino, Leo L. and Lee, Yang W. and Wang, Richard Y.},
  year = {2002},
  month = apr,
  journal = {Commun ACM},
  publisher = {ACM},
  volume = {45},
  number = {4},
  pages = {211–218},
  issn = {0001-0782},
}

@article{batini2009methodologies,
  title = {Methodologies for data quality assessment and improvement},
  author = {Batini, Carlo and Cappiello, Cinzia and Francalanci, Chiara and Maurino, Andrea},
  year = {2009},
  month = jul,
  journal = {ACM Comput Surv},
  publisher = {ACM},
  volume = {41},
  number = {3},
  articleno = {16},
}

@article{rennels2023domain,
	title        = {{How domain experts use an embedded DSL}},
	author       = {Rennels, Lisa and Chasins, Sarah E},
	year         = 2023,
	journal      = {Proc ACM Program Lang},
	publisher    = {ACM},
	volume       = 7,
	number       = {OOPSLA2},
	pages        = {1499--1530}
}

@Article{lehner2025model,
  author    = {Lehner, Daniel and others},
  title     = {{Model-Driven Engineering for Digital Twins: A Systematic Mapping Study}},
  pages     = {1--39},
  journal   = {{Soft Sys Mod}},
  publisher = {{Springer}},
  year      = {2025},
}

@Article{zech2026model,
  author    = {Zech, Philipp and others},
  title     = {{Model-based digital twin engineering: insights, challenges, and future directions}},
  pages     = {1--43},
  journal   = {{Soft Sys Mod}},
  publisher = {{Springer}},
  year      = {2026},
}

@article{whittle2013state,
  title={{The State of Practice in Model-driven Engineering}},
  author={Whittle, Jon and Hutchinson, John and Rouncefield, Mark},
  journal={IEEE software},
  volume={31},
  number={3},
  pages={79--85},
  year={2013},
  publisher={IEEE}
}

@inproceedings{david2024interoperability,
  title = {Interoperability of Digital Twins: Challenges, Success Factors, and Future Research Directions},
  author = {David, Istvan and Shao, Guodong and Gomes, Claudio and Tilbury, Dawn and Zarkout, Bassam},
  year = {2024},
  booktitle = {Leveraging Applications of Formal Methods, Verification and Validation. Application Areas},
  publisher = {Springer},
  pages = {27--46},
}
